\documentclass[11pt]{article} 
\newif\ifpdf
\ifx\pdfoutput\undefined
\pdffalse 
\else
\pdfoutput=1 
\pdftrue
\fi

\ifpdf
\usepackage[pdftex]{graphicx}
\else
\usepackage{graphicx}
\fi

\newcommand{\D}{\Delta}

\begin{document} \title{\bf From  Heisenberg to G\" odel  via
Chaitin\footnote{Partially supported by the  Vice-Chancellor's
University Development Fund 23124.}\phantom{x}\footnote{\bf This paper was 
published without the permission of the authors, without being proof-read,
and erroneously included Dr. Svozil (through no fault of his own) on this 
and five other papers in:
{\em International Journal of Theoretical Physics} 44, 7 (2005),
1053--1065.}\\[-13ex] \hfill\mbox{\it
{\small In mathematics you don't understand things. }}\\[-1ex]
\hfill\mbox{\it {\small You  just get used to them.\phantom{x}}}{\rm
{\small J. von Neumann}\phantom{x}}\\[13ex]} \author{Cristian S. Calude
and Michael A. Stay\\ Department of Computer Science\\ The University of
Auckland\\ New Zealand\\ Email: {\tt
$\{$cristian,msta039$\}$@ec.auckland.ac.nz}}

\date{}

\maketitle

\begin{abstract} In 1927 Heisenberg discovered that the ``more precisely
the position is determined, the less precisely the momentum is known in
this instant, and vice versa''. Four years later G\"odel showed that a 
finitely specified, consistent formal system which is large enough to
include arithmetic is incomplete. As both results express some kind of
impossibility it is natural to ask whether there is any relation between
them, and, indeed, this question has been repeatedly asked for a long
time. The main interest seems to have been in possible implications of
incompleteness to physics.  In this note we  will take interest in the
{\it converse} implication and will offer a positive answer to the
question: Does uncertainty imply incompleteness? We will show that
algorithmic randomness  is equivalent to a ``formal uncertainty
principle'' which implies Chaitin's information-theoretic
incompleteness.  We also show that the derived uncertainty relation, for
many computers, is physical. In fact, the formal uncertainty principle
applies to {\it all} systems governed by the wave equation, not just
quantum waves. This fact supports the conjecture that  uncertainty
implies algorithmic randomness not only in mathematics, but also in
physics.

\end{abstract} \thispagestyle{empty} \section{Introduction} Are there
any connections between uncertainty and incompleteness? We don't know of
any reaction of Heisenberg to this question. However, G\"odel's
hostility to any suggestion regarding possible connections between  his
incompleteness theorem and physics, particularly, Heisenberg's
uncertainty relation, is well-known.\footnote{J. Wheeler was thrown out
of G\"odel's office for asking the question ``Professor G\"odel, what
connection do you see between your incompleteness theorem and 
Heisenberg's uncertainty principle?'', cf. Chaitin's account cited in
Barrow \cite{barrow}, p. 221.}  One of the obstacles in establishing
such a connection comes from the different nature of these two results:
uncertainty is a quantitative phenomenon while incompleteness is
prevalently qualitative.

In recent years there have been a lot of interest in the relations
between computability and incompleteness and physics. Opinions vary
considerably, from the conclusion  that the impact on G\"odel and Turing
incompleteness theorems to physics is a red herring (see
\cite{casti1,casti2}), to Hawking's view that ``a physical theory is
self-referencing, like in G\"odel's theorem. \ldots \, Theories we have
so far are both inconsistent and incomplete'' (cf. \cite{Hawking}). A
very interesting analysis of the possible impact of G\"odel's
incompleteness theorems in physics was written by Barrow
\cite{barrow,barrow0}; the prevalence of physics over mathematics is
argued by Deutsch  \cite{deutsch-97}; for Svozil \cite{karl,karlc},
Heisenberg's incompleteness is pre-G\"odelian-Turing and finite. Other
relevant papers are Geroch and Hartle \cite{gh}, Peres \cite{Peres}, and
Peres and Zurek \cite{PeresZurek}.

In this note we do {\it not} ask whether G\"odel's incompleteness has
any bearing on Heisenberg's uncertainty, but the converse: Does
uncertainty imply incompleteness? We will show that we can get a
positive answer to this question: algorithmic randomness can  be recast
as a ``formal uncertainty principle'' which implies Chaitin's
information-theoretic version of G\"odel's  incompleteness.

\section{Outline} We begin with overviews of the relevant ideas first
discovered by Heisenberg, G\"odel, and Chaitin.

Next, we show that random reals, of which Chaitin Omega numbers are just
an example, 
satisfy  a ``formal uncertainty principle'', namely 
\begin{equation}
\label{0u} \Delta_s\cdot \Delta_{C}(\omega_1\ldots\omega_s) \ge
\varepsilon, 
 \end{equation} where
$\varepsilon$ is a fixed positive constant.

The two conjugate coordinates are the random real  and the binary
numbers describing the programs that generate its prefixes.  Then, the
uncertainty in the random real given an $n$-bit prefix is $2^{-n}$, and
the uncertainty in the size of the shortest program that generates it
is, to within a multiplicative constant, $2^n$.

The Fourier transform is a lossless transformation, so all the
information contained in the delta function $\delta_{\Omega(x)} = 1$ if
$x = \Omega$, $\delta_{\Omega(x)} = 0$,  otherwise, is preserved in the
conjugate. Therefore,  if you need $n$ bits of information to describe a
square wave convergent on the delta function,  there must be $ n$  bits
of information in the Fourier transform of the square wave.  Since both
the information in the transformed square wave and the shortest program
describing the square wave increase linearly with $n$, there is an
equivalence between the two.

 We show that the formal uncertainty principle is a true
uncertainty principle--that is, the terms are bounded by the standard
deviations of two random variables with particular probability
distributions.  We note that for many self-delimiting Turing machines
$C$, the halting probability $\Omega_C$ is computable; in these cases,
there are quantum systems with observables described by these
probability distributions, and our uncertainty relation is equivalent to
Heisenberg's.

Finally,  (\ref{0u}) implies a strong version of G\"odel's
incompleteness, Chaitin's information-theoretic version
\cite{chaitin75,ch75} (see also the analysis in
\cite{jeanpaul,cris2002}). Chaitin's proof relied on measure theory; we
present here a new proof via a complexity-theoretic argument.

\section{Heisenberg} In 1925 Heisenberg developed the theory of matrix
mechanics; it was his opinion that only observable quantities should
play any role in a theory.  At the time, all observations came in the
form of spectral absorption and emission lines.  Heisenberg, therefore,
considered the ``transition quantities'' governing the jumps between
energy states to be the fundamental concepts of his theory.  Together
with Born, who realized Heisenberg's transition rules obeyed the rules
of matrix calculus, he developed his ideas into a theory that predicted
nearly all the experimental evidence available.

The next year, Schr\"odinger introduced what became known as wave
mechanics, together with a proof that the two theories were equivalent. 
Schr\"odinger argued that his version of quantum mechanics was better in
that one could visualize the behavior of the electrons in the atom. 
Many other physicists agreed with him.

Schr\"odinger's approach disgusted Heisenberg; in a letter to Pauli (see
\cite{Pauli}), he called Schr\"odinger's interpretation ``crap". 
Publicly, however, he was more restrained.  In \cite{Heisenberg26} he
argued that while matrix mechanics was hard to visualize,
Schr\"odinger's interpretation of wave mechanics was self-contradictory,
and concluded that something was still missing from the interpretation
of quantum theory.

In 1927 Heisenberg published ``\"{U}ber den Anschaulichen Inhalt der
Quantentheoretischen Kinematik und Mechanik'' (see \cite{Heisenberg}) to
provide the missing piece. First, he gave his own definition of
visualization:  ``We believe we have gained intuitive understanding of a
physical theory, if in all simple cases, we can grasp the experimental
consequences qualitatively and see that the theory does not lead to any
contradictions.''  In this sense, matrix mechanics was just as intuitive
as wave mechanics. Next, he argued that terms like ``the position of a
particle'' can only make sense in terms of the experiment that measures
them.

To illustrate, he considered the measurement of an electron by a
microscope.\footnote{Heisenberg might have been so concerned with
uncertainty because in 1923 he almost failed his Ph.D. exam when
Sommerfeld asked  about  (optical) limitations to the resolution of the
microscope.} The accuracy is limited by the wavelength of the light
illuminating the electron; one can use as short a wavelength as one
wishes, but for very short wavelengths, the Compton effect is
non-negligible.  He wrote, (see \cite{Heisenberg}, p.174--175),
\begin{quote} At the instant of time when the position is determined,
that is, at the instant when the photon is scattered by the electron,
the electron undergoes a discontinuous change in momentum. This change
is the greater the smaller the wavelength of the light employed, i.e.,
the more exact the determination of the position. At the instant at
which the position of the electron is known, its momentum therefore can
be known only up to magnitudes which correspond to that discontinuous
change; thus, the more precisely the position is determined, the less
precisely the momentum is known, and conversely. \end{quote} Heisenberg
estimated the uncertainty to be on the order $$\delta_p\cdot \delta_q
\sim \hbar,$$ where $\hbar$ is Planck's constant over $2\pi$.

Kennard (see \cite{kennard}) was the first to publish the uncertainty
relation in its exact form.  He proved in 1927 that for all normalized
state vectors $|\Psi\rangle$, $$\Delta_p\cdot  \Delta_q \ge \hbar/2,$$
where $\Delta_p$ and $\Delta_q$ are standard deviations of momentum and
position, i.e. $$\Delta_p^2 = \langle\Psi|p^2|\Psi\rangle -
\langle\Psi|p|\Psi\rangle ^2; \, \Delta_q^2 =
\langle\Psi|q^2|\Psi\rangle - \langle\Psi|q|\Psi\rangle ^2.$$ Thus,
assuming quantum mechanics is an accurate description of reality, the
formalism is compatible with Heisenberg's principle.

\section{G\"odel} In 1931 G\"odel 
published his (first) incompleteness theorem in \cite{Godel} (see also
\cite{feferman1,feferman2}).  According to the current terminology, he
showed that {\it every formal system which is (1) finitely specified,
(2) rich enough to include the arithmetic, and (3) consistent, is
incomplete}.  That is, there exists an arithmetical statement which  (A)
can be expressed in the formal system, (B) is  true, but (C) is
unprovable within the formal system.

All conditions are necessary. Condition (1) says that there is an
algorithm listing all axioms and inference rules (which could be
infinite). Taking as axioms all true arithmetical statements will not
do, as this set is not finitely listable. A ``true arithmetical
statement" is a statement about non-negative integers which cannot be
invalidated by finding any combination of non-negative integers that
contradicts it. Condition (2) says that the formal systems has all the
symbols and axioms used in arithmetic, the symbols for  $0$ (zero), $S$
(successor), $+$ (plus),  $\times$ (times), $=$ (equality) and the
axioms making them work (as for example, $x +S(y) = S(x+y)$). Condition
(2) cannot be satisfied if you do not have individual terms for $0, 1,
2, \dots $; for example, Tarski \cite{Tarski} proved that the plane
Euclidean geometry, which refers to points, circles and lines, is
complete.\footnote{This result combined with with G\"odel's completeness
theorem implies  decidability: there is an algorithm which accepts as
input an arbitrary statement of plane Euclidean geometry, and outputs
``true'' if the statement is true, and ``false'' if it is false. The
contrast between the completeness of plane Euclidean geometry and the
incompleteness of  arithmetic is striking.} Finally (3) means that the
formal system is free of contradictions.

Like uncertainty, incompleteness has provoked a lot of interest (and
abuse).

\section{Chaitin} Chaitin has obtained three types of
information-theoretic incompleteness results (scattered through
different publications, \cite{chaitin75,ch75,ch82,chaitin3}; see also
\cite{ch99,ch02}). The strongest form concerns the computation of the
bits of a Chaitin Omega number $\Omega_{U}$, the halting probability of
a self-delimiting universal Turing machine $U$ (see also the analysis in
\cite{jeanpaul,cris2002}). A self-delimiting  Turing machine $C$ is  a
normal Turing machine $C$  which processes binary strings into binary
strings and has a prefix-free  domain, that is, if $C(x)$ is defined and
$y$ is either a proper prefix or an extension of $x$, then $C(y)$  is
not defined. The self-delimiting  Turing machine $U$ is universal if for
every self-delimiting  Turing machine $C$ there exists a fixed binary
string $p$ (the simulator) such that for every input $x$, $U(px) =
C(x)$: either both computations $U(px)$ and $C(x)$ stop and, in this
case they produce the same output or both computations never stop. The
Omega number introduced in \cite{chaitin75} \begin{equation}
\label{omegadef} \Omega_{U} = 0.\omega_{1}\omega_{2}\ldots
\omega_{n}\ldots \end{equation} is the halting probability of $U$; it is
one of the most important concepts in algorithmic information theory
(see \cite{cris2002}).

In \cite{chaitin75} Chaitin  proved the following result: {\it Assume
that $X$ is a formal system satisfying conditions (1), (2) and (3) in
G\"odel's incompleteness theorem. Then, for every   self-delimiting
universal Turing machine $U$, $X$  can determine the positions
and values of only finitely scattered bits of $\Omega_U$, and one can
give a bound on the number of bits of $\Omega_U$ which $X$ can
determine.} This is a form of incompleteness because, with the exception
of finitely many $n$, any true statement of the form ``the $n$th bit of
$\Omega_{U}$ is $\omega_{n}$'' is unprovable in $X$.

For example, we can take $X$ to be $ZFC$\footnote{Zermelo-Fraenkel set
theory with choice.} under the assumption that it is arithmetically
sound, that is, any theorem of arithmetic proved by $ZFC$ is
\emph{true}. Solovay \cite{solovay2k} {\it has constructed a specific 
self-delimiting universal Turing machine $S$ (called Solovay machine)
such that $ZFC$ cannot determine any bit of $\Omega_S$}. In this way one
can obtain {\it constructive} versions of Chaitin's theorem. For
example, {\it if $ZFC$ is arithmetically sound and $S$ is a Solovay
machine, then the  statement ``the  $0$th bit of the binary expansion of
 $\Omega_S $  is  $0$" is true but unprovable in $ZFC$.} In fact, one
can effectively construct arbitrarily many examples of true and
unprovable statements of the above form, cf. \cite{crisomega}.

\section{Rudiments of Algorithmic Information Theory} In this section we
will present some basic facts of algorithmic information theory in a
slightly different form which is suitable for the results appearing in
the following section.

We will work with binary strings; the length of the string $x$ is
denoted by $|x|$. For every $n\ge 0$ we denote by $B(n)$ the binary
representation of the number $n+1$ without the leading 1. For example,
$0 \,\, \mapsto \,\, \lambda$ (the empty string), $1 \,\, \mapsto \,\,
0$, $2 \,\, \mapsto \,\, 1$, $3 \,\, \mapsto \,\, 00, \ldots$ The length
of $B(n)$ is almost equal to $\log_2 (n)$; more precisely, it is
$\lfloor \log_2 (n+1) \rfloor$. The function $B$ is bijective and we
denote by $N$ its inverse. The string $x$ is length-lexicographically
less than the string $y$ if and only if $N(x) < N(y)$.

We need first the Kraft-Chaitin theorem: {\it Let $n_{1}, n_{2}, \ldots
$ be a computable sequence of non-negative integers such that}
\begin{equation} \label{kraft} \sum_{i=1}^{\infty} 2^{-n_{i}} \le 1.
\end{equation} {\it Then, we can effectively construct a prefix-free
sequence of strings (that is, no $w_{i}$ is a proper prefix of any
$w_{j}$ with $i\not= j$) $w_{1}, w_{2}, \ldots $ such that for each
$i\ge 1, \, |w_{i}|=n_{i}$.}

Let $C$ be a self-delimiting  Turing machine. The program-size
complexity induced by $C$ is defined by $H_{C}(x) = \min\{ |w|\mid
C(w)=x\}$ (with the convention that strings not produced by $C$ have
infinite complexity). One might suppose that the complexity of a string
would vary greatly between choices of self-delimiting Turing machine. 
However, because of the universality requirement, the complexity
difference between $C$ and $C'$ is at most the length of the shortest
program for $C'$ that simulates $C$.  Therefore, the complexity of a
string is fixed to within an additive constant.  This is known as the
``invariance theorem'' (see \cite{cris2002}), and is usually stated:
{\it For every self-delimiting universal Turing machine $U$ and
self-delimiting Turing machine $C$ there exists a constant $\varepsilon
>0$ (which depends upon $U$ and $C$) such that for every string $x$,}
\[H_{U}(x) \le \varepsilon + H_{C}(x).\]

For our aim it is more convenient to define the complexity measure
$\nabla_{C}(x) = \min\{ N(w)\mid C(w)=x\}$, the smallest integer whose
binary representation produces $x$ via $C$. Clearly, for every string
$x$, \[2^{H_{C}(x)}-1 \le \nabla_{C}(x) < 2^{H_{C}(x)+1}-1.\] Therefore
we can say that $\Delta_C(x)$, our uncertainty in the value
$\nabla_C(x)$, is the difference between the upper and lower bounds
given, namely $\Delta_C(x) = 2^{H_{C}(x)}$.

The invariance theorem can now be stated as follows: {\it for every
self-delimiting universal Turing machine $U$ and self-delimiting Turing
machine $C$ there exists a constant $\varepsilon >0$ (which depends upon
$U$ and $C$) such that for every string $x$,} \[\D_{U}(x) \le
\varepsilon \cdot \D_{C}(x).\] Let $\Delta_{s} = 2^{-s}$. Chaitin's
theorem (see \cite{chaitin75}) stating that the bits of $\Omega_{U}$ in
(\ref{omegadef}) form a random sequence can now be presented as a 
``formal uncertainty principle'': {\it for every self-delimiting  Turing
machine $C$ there is a constant $\varepsilon >0$ (which depends upon $U$
and $C$) such that} \begin{equation} \label{u} \D_{s} \cdot
\D_{C}(\omega_{1}\ldots \omega_{s}) \ge \varepsilon. \end{equation}

The inequality (\ref{u}) is an uncertainty relation as it reflects a
limit to which we can simultaneously increase both the accuracy with
which we can approximate $\Omega_{U}$ and the complexity of the initial
sequence of bits we compute; it relates the uncertainty of the output to
the size of the input. When $s$ grows indefinitely, $\D_{s} $ tends to
zero, in contrast with $\D_{C}(\omega_{1}\ldots \omega_{s})$ which tends
to infinity; their product is not only bounded from below, but increases
indefinitely (see also (\ref{uu})). From a complexity viewpoint
(\ref{u}) tells us that there is a limit  $\varepsilon$  up to which we
can uniformly compress the initial prefixes of the binary expansion  of 
$\Omega_{U}$.

How large can be $\varepsilon$ in (\ref{u})?  For example, $\varepsilon
= 1$  when  $C=U_{0}$ is a special universal self-delimiting  Turing
machine: \begin{equation} \label{u1} \D_{s} \cdot
\D_{U_{0}}(\omega_{1}\ldots \omega_{s}) \ge 
 1.
\end{equation} If $U$ is universal and satisfies (\ref{u}), then  a
universal machine $U_{0}$ satisfying (\ref{u1}) can be  defined by
$U_{0}(0^{\varepsilon}x) = U(x)$ (so requiring that any input to $U_0$
not starting with $\varepsilon$ zeros causes the machine to go into an
infinite loop).

In fact, in view of the strong complexity-theoretic characterization of
random sequences (see \cite{chaitin75,cris2002}) a stronger form of
(\ref{u}) is true: {\it for every positive integer $N$ there is a bound
$M$ (which depends upon $U$, $C$ and $N$) such that for all $s \ge M$ we
have:} \begin{equation} \label{uu} \D_{s} \cdot \D_{C}(\omega_{1}\ldots
\omega_{s}) \ge N. \end{equation} The constant $N$ appearing in
(\ref{u}) can be made arbitrarily large in case $s$ is large enough; the
price paid appears in the possible violation of the inequality for the
first $s <M$ bits.

\if01 Is (\ref{u}) a `true' uncertainty relation? We prove that the
variables $\Delta_{s}$ and $\Delta_{C}$ in (\ref{u}) are  the standard
deviations of two measurable observables in suitable probability 
spaces.

For $\Delta_{s}$ we consider the space of all real numbers in the unit
interval which are approximated to exactly $s$ digits. Fix the first $s$
digits of $\Omega_{U}$, $\omega_{1}\omega_{2}\ldots \omega_{s}$ and
define the random variable $X$ on a real approximated by its first $s$
digits $v = v_{1}v_{2}\ldots v_{s}$ by the delta function $X(v)=1$
(v_{s}\oplus\omega_{s}). if $v = \omega_{1}\omega_{2}\ldots \omega_{s}$
and $X(v) =0$ otherwise. Each approximation is equally probable, so
$P(v)= 2^{-s}$.  Then the expectation values of both $X$ and $X^2$ are
$\langle X^2\rangle = \langle X\rangle = 2^{-s}$ and the standard
deviation is $\sigma_X = 2^{-s/2}\cdot (1-2^{-s})^{1/2}\ge c_{1}\cdot
\Delta_{s}$. 
(2^{s}-1)\cdot2^{s+1})$ Moreover, when $s \to \infty$, both $ \sigma_{X}
$ and $\Delta_{s}$ tend to zero.

For $\Delta_{C}$ we consider the same space but the  random variable $Y
(v_{1} v_{2}\ldots v_{s}) =
0.(v_{1}\oplus\omega_{1})(v_{2}\oplus\omega_{2})\ldots
(v_{s}\oplus\omega_{s})$ (where $\oplus$ is the sum modulo 2) and the
probability is $Prob(v) = P_{C}(v)/\Omega_{C}^{s}$,  where $P_{C}(x) =
\sum_{C(y) =x} 2^{-|y|}$ and $\Omega_{C}^{s}= \sum_{|x| =s} P_{C}(x)$.
Let $m = \min_{|x|=s} P_{C}(x)$ and $M = \max_{|x|=s} P_{C}(x)$. Then,
\[\langle Y\rangle \le \frac{M}{\Omega_{C}^{s}}\cdot \sum_{|v|=s} Y(v) =
\frac{M}{2\cdot \Omega_{C}^{s}}\cdot (2^{s}-1),\] \[\langle Y^2\rangle
\ge \frac{m}{6\cdot \Omega_{C}^{s}} \cdot (2^{s}-1)(2^{s+1}+1),\]
\[\sigma_{Y}^{2} \ge \frac{m}{6\cdot \Omega_{C}^{s}} \cdot
(2^{s}-1)(2^{s+1}+1) - \frac{M}{2 \cdot \Omega_{C}^{s}} \cdot
(2^{s}-1),\] hence we get: \begin{equation} \label{us} \sigma_Y \ge
c_{2}\cdot \Delta_{C}(\omega_{1}\omega_{2}\ldots \omega_{s}).
\end{equation}

Again, when $s \to \infty$, both $\sigma_Y$ and $\Delta_{C}
(\omega_{1}\omega_{2}\ldots \omega_{s})$ tend to infinity.

Putting the  inequalities (\ref{u})  and (\ref{us}) together, we have
that $$\sigma_X \cdot \sigma_Y \ge c_1 \cdot c_2 \cdot \Delta_s \cdot
\Delta_{C}(\omega_{1}\omega_{2}\ldots \omega_{s}) \ge c_1 \cdot c_2
\cdot \varepsilon,$$ so for $U_{0}$ satisfying  (\ref{u1}) we have:
\[\sigma_X \cdot \sigma_Y \ge c_1 \cdot c_2.\] It is worth noticing that
$\sigma_{X}\cdot \sigma_{Y}$ tends to infinity when $s\to\infty$, in
accord with the stronger relation (\ref{uu}). \fi

Is (\ref{u}) a `true' uncertainty relation? We prove that the variables
$\Delta_{s}$ and $\Delta_{C}$ in (\ref{u}) are  standard deviations of
two measurable observables in suitable probability spaces.

For $\Delta_{s}$ we consider the space of all real numbers in the unit
interval which are approximated to exactly $s$ digits. Consider the
probability distribution $Prob(v) = P_{C}(v)/\Omega_{C}^{s}$,  where
$P_{C}(x) = \sum_{C(y) =x} 2^{-|y|}$ and $\Omega_{C}^{s}= \sum_{|x| =s}
P_{C}(x)$.

Now fix the first $s$ digits of $\Omega_{U}$,
$\omega_{1}\omega_{2}\ldots \omega_{s}$ and define $$\alpha =
2^{-s/2}\cdot (Prob(\omega_{1}\omega_{2}\ldots \omega_{s}))^{-1/2} \cdot
(1- Prob(\omega_{1}\omega_{2}\ldots \omega_{s}))^{-1/2}.$$ The random
variable $X$ on a real approximated by the first $s$ digits $v =
v_{1}v_{2}\ldots v_{s}$ is defined by the delta function $X(v)= \alpha$
if $v=\omega_{1}\omega_{2}\ldots \omega_{s}$ and $X(v) =0$ otherwise.
Then the expectation values of $X$ and $X^2$ are $\langle X\rangle =
\alpha \cdot Prob(\omega_{1}\omega_{2}\ldots \omega_{s})$ and $\langle
X^2\rangle = \alpha^{2} \cdot Prob(\omega_{1}\omega_{2}\ldots
\omega_{s})$, so the standard deviation is $\sigma_X =  \Delta_{s}$.

For $\Delta_{C}$ we consider $$\beta =
(\D_{C}(\omega_{1}\omega_{2}\ldots \omega_{s}))^{1/2}\cdot
(Prob(\omega_{1}\omega_{2}\ldots \omega_{s}))^{-1/2}\cdot (1-
Prob(\omega_{1}\omega_{2}\ldots \omega_{s}))^{-1/2},$$ and the same
space but the  random variable $Y (\omega_{1}\omega_{2}\ldots
\omega_{s}) = \beta$ and  $Y(v) =0$ if $v \not=
\omega_{1}\omega_{2}\ldots \omega_{s}$. Then, the expectation values of 
$Y$ and $Y^2$ are $\langle Y\rangle = \beta \cdot
Prob(\omega_{1}\omega_{2}\ldots \omega_{s})$ and $\langle Y^{2}\rangle =
\beta^{2} \cdot Prob(\omega_{1}\omega_{2}\ldots \omega_{s})$, so the
standard deviation is $\sigma_Y = \D_{C}(\omega_{1}\omega_{2}\ldots
\omega_{s})$.

Hence the relation (\ref{u}) becomes:

$$\sigma_X \cdot \sigma_Y = \Delta_s \cdot
\Delta_{C}(\omega_{1}\omega_{2}\ldots \omega_{s}) \ge  \varepsilon,$$ so
for $U_{0}$ satisfying  (\ref{u1}) we have: \[\sigma_X \cdot \sigma_Y
\ge 1.\]

\section{From Heisenberg to Chaitin}

Since self-delimiting universal Turing machines are strictly more
powerful than non-universal ones, the inequality holds for the weaker
computers as well. In many of these cases, the halting probability of
the machine is computable, and we can construct a quantum algorithm to
produce a set of qubits whose state is described by the distribution.

To illustrate, we consider a quantum algorithm with two parameters, $C$
and $s$, where $C$ is a Turing machine for which the probability of
producing each $s$-bit string is computable.  We run the algorithm to
compute that distribution on a quantum computer with $s$ ouput qubits;
it puts the output register into a superposition of spin states, where
the probability of each state $|v \rangle$ is $P_{C}(v)/\Omega_{C}^{s}$.
Next, we apply the Hamiltonian operator $H=\beta|\omega_1\ldots\omega_s
\rangle \langle \omega_1\ldots\omega_s|$ to the prepared state. A
measurement of energy will give $\beta$ with probability
$P=Prob(\omega_{1}\omega_{2}\ldots \omega_{s})$ and zero with
probability $1-P$. The expectation value for energy, therefore, is
exactly the same as that of $Y$, but with units of energy, i.e.
$$\Delta_C(\omega_{1}\omega_{2}\ldots \omega_{s}) [J] \cdot \Delta_s \ge
\varepsilon [J],$$ where $[J]$ indicates Joules of energy.

Now define $$\Delta_t \equiv \frac{\sigma_Q}{|d\langle
Q\rangle/dt|}\raisebox{.5ex}{,}$$ where $Q$ is any observable that does
not commute with the Hamiltonian; that is, $\Delta_t$ is the time it
takes for the expectation value of $Q$ to change by one standard
deviation.  With this definition, the following is a form of
Heisenberg's uncertainty principle: $$\Delta_E \cdot \Delta_t \ge
\hbar/2.$$ We can replace $\Delta_E$ by
$\Delta_C(\omega_{1}\omega_{2}\ldots \omega_{s})$ by the analysis above;
but what about $\Delta_t$? If we choose a time scale such that our two
uncertainty relations are equivalent for a single quantum system
corresponding to a computer $C$ and {\it one} value of $s$, then the
relation holds for $C$ and {\it any} value of $s$:
$$\Delta_C(\omega_{1}\omega_{2}\ldots \omega_{s}) [J] \cdot \Delta_s
\frac{\hbar}{2\varepsilon} [J^{-1} \cdot Js] \ge \frac{\hbar}{2} [Js].$$
In this sense, we claim that Heisenberg's uncertainty relation is
equivalent to (\ref{u}).  We cannot say whether (\ref{u}) is physical
for universal self-delimiting Turing machines; to do so requires
deciding the Church-Turing thesis for quantum systems.

The uncertainty principle now says that getting one more bit of
$\Omega_U$ requires (asymptotically) twice as much energy.  Note,
however, that we have made an arbitrary choice to identify energy with
complexity.  We could have chosen to create a system in which the
position of a particle corresponded to the complexity, while momentum
corresponded to the accuracy of $C$'s estimate of $\Omega_U$.  In that
case, the uncertainty in the position would double for each extra bit. 
Any observable can play either role, with a suitable choice of units.

If this were the only physical connection, one could argue that the
result is merely an analogy and nothing more.  However, consider the
following:  let $\rho$ be the density matrix of a quantum state.  Let
$R$ be a computable positive operator-valued measure, defined on a
finite dimensional quantum system, whose elements are each labeled by a
finite binary string. Then the statistics of outcomes in the quantum
measurement is described by $R$:  $R(\omega_{1}\ldots \omega_{s})$ is
the measurement outcome and $tr(\rho R(\omega_{1}\ldots \omega_{s}))$ is
the probability of getting that outcome when we measure $\rho$. Under
these hypotheses, Tadaki's inequality (1) (see \cite{tadaki}, p. 2), and
our inequality  (\ref{u}) imply the existence of a constant $\tau$
(depending upon $R$) such that for  all $\rho$ and $s$ we have: \[
\Delta_{s}\,  \raisebox{.6ex}{.}\, \frac{1}{tr(\rho R(\omega_{1}\ldots
\omega_{s}))} \ge \tau.\] In other words, there is no algorithm that,
for all $s$, can produce \begin{enumerate} \item an experimental setup
to produce a quantum state and \item a POVM with which to measure the
state such that \item the probability of getting the result
$\omega_{1}\omega_{2}\ldots \omega_{s}$ is greater than $1/(\tau 2^s)$.
\end{enumerate}

Finally, it is interesting to note that a Fourier transform of the wave
function switches between an ``Omega space'' and a ``complexity space''.
We plan on examining this relationship further in a future paper.

\section{From Chaitin to G\" odel} In this section we prove that the
uncertainty relation (\ref{u}) implies incompleteness.

We start with the following theorem: {\it Fix a universal
self-delimiting Turing machine $U$. Let $x_{1}x_{2}\ldots $ be a binary
infinite sequence and let $F$ be a strictly increasing function mapping
positive integers into positive integers. If the set $\{(F(i),
x_{F(i)})\mid i \ge 1\}$ is computable, then there exists a constant
$\varepsilon >0$ (which depends upon $U$ and the characteristic function
of the above set) such that for all $k\ge 1$ we have:} \begin{equation}
\label{bc} \Delta_{U}(x_{1}x_{2}\ldots x_{F(k)}) \le \varepsilon \cdot
2^{F(k)-k}. \end{equation} To prove (\ref{bc}) we consider for  every
$k\ge 1$ the strings \begin{equation} \label{s}
w_{1}x_{F(1)}w_{2}x_{F(2)}\ldots w_{k}x_{F(k)}, \end{equation} where
each $w_{j}$ is a string of length $F(j)-F(j-1)-1, F(0)=0$, that is, all
binary strings of length $F(k)$ where we have fixed  bits at the
positions $F(1), \ldots ,F(k)$.

It is clear that $\sum_{i=1}^{k} |w_{i}| = F(k)-k$ and the mapping
$(w_{1}, w_{2},  \ldots ,w_{k}) \mapsto w_{1}w_{2} \ldots w_{k}$ is
bijective, hence to generate all strings of the form (\ref{s}) we only
need to generate all strings of length $F(k)-k$.

Next we consider the enumeration of all strings of the form (\ref{s})
for $k=1,2,\ldots$. The lengths of these strings will form the sequence
\[ \underbrace{F(1), F(1), \ldots ,F(1)}_{2^{F(1)-1}\mbox{ times}},
\ldots \underbrace{,F(k), F(k), \ldots ,F(k)}_{2^{F(k)-k}\mbox{ 
times}}, \ldots\] which is computable and satisfies the inequality
(\ref{kraft}) as \[\sum_{k=1}^{\infty}  2^{F(k)-k}\cdot 2^{-F(k)} = 1.\]

Hence, by Kraft-Chaitin theorem, for every string $w$ of length $F(k)-k$
there effectively exists a string $z_{w}$ having the same length as $w$
such that the set $\{z_{w}\mid |w|=F(k)-k, k\ge 1\}$ is prefix-free.
Indeed, from  a string $w$ of length $F(k)-k$  we get  a unique
decomposition $w = w_{1}\ldots w_{k}$, and $z_{w}$ as above, so we can
define   $C(z_{w}) = w_{1}x_{F(1)}w_{2}x_{F(2)}\ldots w_{k}x_{F(k)}$;
$C$ is a  self-delimiting Turing machine. Clearly, 
\[\Delta_{C}(w_{1}x_{F(1)}w_{2}x_{F(2)}\ldots w_{k}x_{F(k)}) \le
\nabla_{C}(w_{1}x_{F(1)}w_{2}x_{F(2)}\ldots w_{k}x_{F(k)}) \le N(z_w)
\le 2^{F(k)-k+1}-1,\] for all $k\ge 1$. In particular,
$\Delta_{C}(x_{1}\ldots x_{F(k)}) \le 2^{F(k)-k+1}-1$, so by the
invariance theorem we get the inequality (\ref{bc}).

It is easy to see that under the hypothesis of the above theorem the
uncertainty relation (\ref{u}) is violated, so the sequence
$x_{1}x_{2}\ldots x_{n}\ldots $ is not random. Indeed, if the sequence
were random, then the formal  uncertainty principle (\ref{u}) will hold
true, hence for each $k\ge 1$, we would have the following contradictory
pair of inequalities: \[\varepsilon_{1}\cdot \frac{1}{\Delta_{F(k)}} \le
\Delta_{U} (x_{1}\ldots x_{F(k)}) \le \varepsilon\cdot 2^{F(k)-k}.\]

We are now able to deduce Chaitin's information-theoretic incompleteness
theorem from the uncertainty relation (\ref{u}). Assume by absurdity
that $ZFC$ can determine infinitely many digits of $\Omega_{U} =
0.\omega_{1}\omega_{2}\ldots$. Then, we could enumerate an infinite
sequence of digits of $\Omega_{U}$, thus contradicting the above
theorem.

In particular, there exists a bound $N$ such that $ZFC$ cannot determine
more than $N$ scattered digits of $\Omega_{U} =
0.\omega_{1}\omega_{2}\ldots$.

\section{Conclusion}

We have shown that uncertainty implies algorithmic  randomness which, in
turn, implies incompleteness.  Specifically, the complexity-theoretic
characterization of the randomness of the halting probability of a
universal self-delimiting Turing machine $U$, Chaitin Omega number
$\Omega_U$,  can be recast as a ``formal uncertainty principle'':  an
uncertainty relation between the accuracy of one's estimate of
$\Omega_U$ and the complexity of the initial bit string.  This relation
implies Chaitin's information-theoretic version of G\"odel's 
incompleteness.

The uncertainty relation applies to all self-delimiting Turing machines
$C$.  For the class of  machines whose halting probabilities
$\Omega_{C}$  are computable, we have shown that one can construct a
quantum computer for which the uncertainty relation describes conjugate
observables.  Therefore, in these particular instances, the uncertainty
relation is equivalent to Heisenberg's.

There is an important distinction between ``quantum randomness" and 
our formal  uncertainty principle.  They are separate
concepts.  In the Copenhagen  interpretation, the random collapse of the
wave-function is a postulate.  In the  Bohmian interpretation, where
there are real particles with real (though non-Newtonian) trajectories,
randomness comes from our ignorance about the system; the velocity of
any particle depends instantaneously on every other particle.    In one
case the interpretation is probabilistic, while in the other, it is
completely deterministic.  We cannot distinguish between these. Our
result concerns a different source of randomness.

Like Heisenberg's uncertainty principle, our formal uncertainty principle
is a general one; they both apply to {\it all} systems governed by the
wave equation, not just quantum waves.   We could, for  example, use
sound waves instead of a quantum system by playing two pure tones with
frequencies $f$ and $f+\Delta_C(\omega_1\ldots\omega_s)$.  Then
$\Delta_s$ corresponds to the complementary observable, the length of
time needed to perceive a beat.  The (algorithmic) randomness we are
concerned with seems to be pervasive in physics, even at the classical
level.  We may speculate that uncertainty implies randomness not only in
mathematics, but also in physics.

\section*{Ackowledgement} We thank K. Svozil  for suggesting the
references \cite{casti1,casti2} and speculating that ``uncertainty
implies randomness  in physics''.

 \end{document} \end